\documentstyle[aps]{revtex}
\begin{document}
\draft
\twocolumn[\hsize\textwidth\columnwidth\hsize\csname
@twocolumnfalse\endcsname
\title{\rightline{{\tt (December 1998)}}
\rightline{{\tt UM-P-98/62}}
\rightline{{\tt RCHEP-98/18}}
\ \\
Mirror matter and primordial black holes}
\author{N. F. Bell and R. R. Volkas }
\address{School of Physics\\
Research Centre for High Energy Physics\\
The University of Melbourne\\
Parkville 3052 Australia\\
(n.bell@physics.unimelb.edu.au, r.volkas@physics.unimelb.edu.au)}
\maketitle
\begin{abstract}
A consequence of the evaporation of primordial black holes in the early universe may 
be the generation of mirror matter.  This would have implications with regard to dark 
matter, and the number of light particle species in equilibrium at the time of big bang 
nucleosynthesis.  The possibilities for the production of mirror matter by this mechanism 
are explored.
\end{abstract}
\vskip2pc]

Implicit in most discussions concerning the evaporation of primordial black holes (PBHs) 
is the assumption that the only particles produced are those which interact with the 
particles in the standard model.  However there may exist particles which are singlets under 
the standard model gauge group and whose only interaction with ordinary matter is 
gravitational. It has been pointed out in ref.\cite{lip} that the evaporation of 
primordial black holes is a mechanism by which some of this matter may be produced.

An example is the Exact Parity Model (EPM) of ref.\cite{epm}, which involves introducing a 
mirror sector, designed to restore parity symmetry.  Each ordinary particle is related 
via the nonstandard parity symmetry to a mirror partner.
In this model the ordinary and mirror sectors only interact gravitationally, 
and via mixing between the ordinary and mirror neutrinos, Higgs, and neutral gauge bosons.  
Other models exist where gravitation is the only interaction between the two sectors, such as 
the $E_8 \times E_8'$ shadow matter model discussed in ref.\cite{kolbseckelturner}.

The temperature of thermalised mirror matter at the time of big bang nucleosynthesis (BBN)
must be suitably smaller than the temperature of ordinary matter.  The amount of mirror 
baryonic matter in the universe today, however, may be as large as the amount of ordinary 
baryonic matter, as this is a separate issue which depends upon the size of the mirror baryon 
asymmetry.  Despite this, we shall assume as an initial condition that the universe contains 
no mirror matter, a concrete mechanism for which may be the inflationary scenario discussed 
in Ref.\cite{kolbseckelturner}.  We then explore PBH evaporation as a purely gravitational 
process which may produce mirror matter.
It is important for mirror matter models because even if we start out with no mirror matter, 
and if all non-gravitational interactions between the ordinary and mirror sectors are turned 
off, the fundamental gravitational process of PBH evaporation will still exist.

Primordial black holes can be generically produced in the early universe as a 
result of density fluctuations produced during the inflationary epoch.  The impact 
of these black holes for cosmology depends on their mass spectrum.  PBHs of mass 
$M\simeq 5\times 10^{14}\text{g}$ would be evaporating at present, while those of 
mass $M\simeq 10^{9}\text{g}$ evaporate at the time of BBN.  
As cosmology is fairly well understood during the era between BBN and the present, 
constraints exist, some of which are quite stringent, on the number density of black 
holes which evaporate during this period.  
These include such constraints as limits on the flux of 100MeV gamma rays from evaporations 
at the present time as well as not disrupting standard BBN or destroying the deuterium
abundance after BBN.
A summary of the various constraints may be found in
\cite{carrlidsey,liddlegreen} and references therein.
Less constrained is the density of PBHs which evaporate well before BBN, where it 
may even be possible for PBHs to briefly dominate the energy density of the universe 
before evaporation.  

The current determination of cosmological parameters leaves the number density of 
PBHs quite model dependent and open to various possibilities.  
Hence, the purpose of this paper is not to precisely calculate the mirror matter 
density of the universe, but rather to point out some of the possibilities.

Firstly, what do we consider a significant amount of mirror matter?  The number of 
light degrees of freedom at the time of nucleosynthesis is constrained by the 
successful BBN calculations, and it is standard to parameterise additional degrees 
of freedom in terms of a change in the effective number of neutrinos, 
$\Delta N_{\nu}^{\text{eff}} \equiv N_{\nu}^{\text{eff}}-3$. 
The size of $\Delta N_{\nu}^{\text{eff}}$ is dependent on observational uncertainties 
in the BBN abundances of light elements, with a recent analysis ref.\cite{olive} 
finding the bound can be anything from $ N_{\nu}^{\text{eff}} < 3.3$ to
$N_{\nu}^{\text{eff}} < 6$.  An amount of mirror 
matter equivalent to $\Delta N_{\nu}^{\text{eff}} \simeq 0.3$ would be cosmologically 
significant, especially given that the upcoming MAP and PLANCK satellites are predicted 
to be sensitive to $\Delta N_{\nu}^{\text{eff}} \agt 0.1$ \cite{raffelt}.

We shall consider two mechanisms which may be responsible for black hole formation, 
namely density perturbations produced during inflation and phase transitions.  
The density perturbation picture is useful because the number of PBH produced in 
this way is calculable, whereas although phase transitions in the early universe are to 
be expected, their implications for PBH formation are more uncertain.

In the standard picture of large scale structure formation, small density perturbations, 
which were produced at inflation and exist over a range of scales, evolved into the 
galaxies and clusters we observe today.  These same density perturbations, although 
at a much smaller scale, could also have been responsible for the formation of PBHs.  
It is customary to assume a power law spectrum of Gaussian density perturbations, 
given by $P(k) \propto k^n$, where $n$ is referred to as the spectral index, and
$P(k)=\langle |\delta_k|^2 \rangle$ with $\delta_k$ being a Fourier component of the 
perturbation.  Inflation tends to predict an approximately scale-invariant spectrum 
$n=1$, although both larger and smaller values are obtained in various inflationary 
models \cite{n}.  
If we take this mechanism, and assume the spectral index remains constant across all 
scales (which represents a large number of orders of magnitude), then the density of 
black holes is large enough to be interesting only if we have $n>1$, a so-called blue 
spectrum.

If a density perturbation is of sufficient size, a black hole is formed at the time 
it re-enters the horizon (Hubble distance), and has a mass of order of the horizon mass at 
that time,
\begin{equation}
M_H \simeq 10^{18} \text{g} \left( \frac{10^7 \text{GeV}}{T_r}\right)^2,
\end{equation}
where $T_r$ is the temperature of the (radiation dominated) universe at the time of re-entry.  
We see that light 
black holes are formed earlier, with the lightest PBHs being those formed directly 
after inflation.  The smallest black holes evaporate first, with the
PBH lifetime given by \cite{tevap}
\begin{equation}
\tau_{\text{evap}} \simeq \frac{10^{-26}}{f(M)}  \left(
\frac{M}{1 \text{g}}\right)^3 \text{sec},
\end{equation}
where $f$ depends on the number of particle species which can be emitted, and is 
normalised such that $f=1$ for black holes which emit only massless particles.

The fraction of the density of the universe in black holes of mass $M$ at the time of 
formation is approximately given by \cite{carr}
\begin{equation}
\label{beta}
\beta_i(M) \simeq \sigma(M) \exp\left(\frac{-1}{18\sigma^2(M)}\right),
\end{equation}
where $\sigma(M)$ is the standard deviation of the density fluctuations when they 
re-enter the horizon.
COBE observations determine $\sigma(M)$ to be of order $10^{-4}$ at large scales, 
which may be related to values at smaller scales according to
\cite{liddlegreen}
\begin{equation}
\label{sigma}
\sigma(M) \simeq 9.5\times 10^{-5}(M/10^{56}g)^{(1-n)/4},
\end{equation}
assuming a radiation dominated universe.

The effect of $n>1$ is to enhance the fluctuations on small scales, and in this 
case $\beta$ is a rapidly decreasing function of $M$.
COBE measurements determine $n=1.2 \pm 0.3$ \cite{cobe} at large scales, and the most 
restrictive of the PBH constraints, which concerns the destruction of deuterium after BBN, 
is $n \alt 1.22$ \cite{liddlegreen}.
However, for $n \sim 1.2$, it is possible to produce an interesting amount of mirror 
matter via evaporation of black holes before BBN.

A question which then arises is: what does $n>1$ do to standard large scale 
structure models (which usually assume $n=1$)?  As is well known, the major problem 
facing standard cold dark matter models is the overproduction of power on smaller 
scales with respect to large scales.  One method which has been tried to alleviate this 
is tilting the power spectrum, so that $n<1$. However, this approach has not met with 
great success, and a more favoured model is to take a hot dark matter component 
consisting of massive neutrinos with $\Omega_{\nu} \simeq 0.2$ so that the 
free-streaming of neutrinos reduces the structure on small scales.
It may be possible to reconcile large scale structure models with $n>1$ (which would 
aggravate the problem of overproduction of structure on small scales)  by increasing 
the hot dark matter component, see ref.\cite{nchdm}.  
However, we now show that even if an $n>1$ model could be made to work from a large scale 
structure point of view, the production of an interesting amount of mirror matter through 
PBH evaporation entails an unattractive fine tuning.

So, let us now calculate the mirror matter abundance as a function of n.
Given that the mass spectrum of PBHs is a strongly decreasing 
function of $M$, we shall consider black holes of a single mass which evaporate 
well before BBN.  
For example, assume that the mirror matter contributes an amount equivalent to 
$\Delta N_{\nu}^{\text{eff}}$ to the density at the time of BBN.  The total density 
at BBN would be
\begin{equation}
\rho_{\text{total}} = \left( \frac{43}{4} + \frac{7}{4}\Delta N_{\nu}^{\text{eff}} 
\right) \frac{\pi^2}{30}T^4 = \rho + \rho',
\end{equation}
where $\rho$ is the density due to ordinary photons, electrons and neutrinos, and $\rho'$ is 
the density of mirror particles.  The ratio $ \rho' / \rho_{\text{total}} $ will be equal to 
half the fraction of primordial black holes $\rho_{\text{PBH}} / \rho_{\text{total}}$ at the 
time of evaporation.
If we define $\alpha$ to be the ratio of the density of PBHs to the density of radiation
\begin{equation}
\alpha = \frac{\rho_{\text{PBH}}}{\rho_{\text{total}}-\rho_{\text{PBH}}},
\end{equation}
at the time of evaporation $\alpha$ must have the value, 
\begin{equation}
\alpha_{\text{evap}} = \frac{14 \Delta N_{\nu}^{\text{eff}}}{43-7\Delta N_{\nu}^{\text{eff}}},
\end{equation}
so that for $\Delta N_{\nu}^{\text{eff}} \leq 0.6$, we have the bound 
$\alpha_{\text{evap}} \leq 0.2$. 
From this value of $\alpha_{\text{evap}}$ we can work out the initial 
density of PBHs 
and the corresponding value of the spectral index $n$.   
Note that while the bound on $\alpha_{\text{evap}}$ is sensitive to 
$\Delta N_{\nu}^{\text{eff}}$, the spectral index is not.

As the PBHs are non-relativistic, 
their energy density dilutes less quickly than the radiation background, so the initial 
ratio of PBHs grows according to \cite{liddlegreen}
\begin{equation}
\label{alpha}
\alpha_{\text{evap}} (M) \simeq 3.2 \frac{\beta_i}{1-\beta_i} \frac{M}{m_{\text{pl}}},
\end{equation}
where $m_{\text{pl}}=2 \times 10^{-5}\text{g}$ is the Planck mass.
Let us assume, for the sake of the example, that inflation occurs at the scale of 
approximately $10^{14} \text{GeV}$, so that black holes formed immediately after inflation 
have a mass $M\sim 10^4\text{g}$.  We may determine $\alpha_{\text{evap}}$ as a function of 
the spectral index $n$, using eqns.(\ref{beta}),(\ref{sigma}) and (\ref{alpha}).  
The initial density of PBHs, $\beta_i(M)$, is a sensitive function of $n$ since we are dealing 
with very small scales and hence large wavenumbers.
If we set $n=1.21$, we obtain $\alpha_{\text{evap}} \simeq 0.04$, whereas
for $n=1.22$ the 
universe is dominated by black holes before their evaporation, resulting in 
equal quantities of matter and mirror matter.  It would therefore be necessary to 
finely tune the spectral index if one wished to obtain a significant amount of mirror 
matter without overproducing it .

Of course, the default assumption, that the spectral index is constant
over all scales may not in fact be the case, noting that
PBHs and structure at the size of clusters of galaxies, correspond to
very different scales.

An extended spectrum of density perturbations is not the only mechanism
which may produce PBHs.
For example, ref.\cite{pt} discusses the case of  two-stage inflation 
whereby a phase transition between the two inflationary stages leads to a
spike in the spectrum of density perturbations,
resulting in black hole formation over a small mass range, where perhaps
the universe becomes black hole dominated before reheating proceeds via
black hole evaporation.
Collapsing cosmic string loops have been considered in ref.\cite{strings}
as a process that may lead to black hole formation.

Phase transitions have been studied in ref.\cite{bubbles,twobubble} as a 
possible mechanism for the production of black holes.
It is natural to expect phase transitions to occur due to gauge
symmetry breaking at temperatures much larger than the electroweak
scale, for example GUT symmetries, left-right symmetries, etc. 
A phase transition must be first order to achieve PBH production, which results 
from the collision of bubbles of true vacuum formed in the 
false vacuum.
Bubble collisions have been studied in ref.\cite{bubbles}, where it was found
that multi-bubble collisions are required with collisions between two bubbles not being 
sufficient to produce black holes.
Ref.\cite{twobubble} however, argues that significant PBHs production can result from 
two-bubble collisions.
As before, the lighter PBHs are those produced at earlier times, but the 
number of black holes produced is uncertain and model dependent.
Any symmetry breaking occurring via a first order phase transition at
an early epoch in the universe could plausibly have produced some black
holes, the evaporation of which could result in a substantial mirror
matter density.

We have been considering the initial fraction of mirror particles in the
radiation dominated universe.
The mirror matter present in the universe today would consist of mirror
photons and neutrinos, so that the mirror particle contribution to the
dark matter would depend on the mass of the mirror neutrinos.
In the EPM, the ordinary and mirror neutrinos of a given generation 
are maximally mixed linear
combinations of two mass eigenstates, with the resolution of the solar and 
atmospheric neutrino anomalies motivating the two mass eigenstates to be 
nearly degenerate.
So, if the neutrinos have a mass such that they constitute a fraction of the density
in the form of hot dark matter, there will also be mirror neutrino component.

In conclusion, if a power law spectrum of density perturbations is responsible for PBH 
formation, their evaporation generically leads to a density of mirror matter 
of either zero or $\rho'=\frac{1}{2}\rho_{\text{total}}$.  Anything in between may be 
obtained with sufficient fine tuning of the spectral index $n$, though we tend to rule 
this out on the grounds of naturalness. This is our main result.
We conclude, therefore, that if a significant amount of mirror matter does exist as a 
result of PBH evaporation, then the black holes are more likely to have been produced 
during phase transitions.

\acknowledgments{RRV is supported by the 
Australian Research Council.  NFB is supported by the 
Commonwealth of Australia and the University of Melbourne.}


\begin{references}

\bibitem{lip}
G.\ E.\ A.\ Matsas, J.\ C.\ Montero, V.\ Pleitez and D.\ A.\ T.\ Vanzella,
Proceedings of Topics in Theoretical Physics II: Festschrift for A.\ H.\ Zimerman, 
Sao Paulo, Brazil, 1998, edited by H.\ Aratyn, J.\ H.\ Ferreira and J.\ F.\ Gomes, p.219,
hep-ph/9810456

\bibitem{epm}
R.\ Foot, H.\ Lew, and R.\ R.\ Volkas, Phys.\ Lett.\ B 272, 67 (1991);
Mod.\ Phys.\ Lett.\ A7,  2567 (1992);
R.\ Foot and R.\ R.\ Volkas, Phys.\ Rev.\ D 52, 6595 (1995), and
references therin.

\bibitem{kolbseckelturner}
E.\ W.\ Kolb, D.\ Seckel and M.\ S.\ Turner, Nature 314, 415 (1985).

\bibitem{carrlidsey}
B.\ J.\ Carr and J.\ E.\ Lidsey, Phys.\ Rev.\ D48, 543 (1993).

\bibitem{liddlegreen}
A.\ M.\ Green and A.\ R.\ Liddle, Phys.\ Rev.\ D56, 6166 (1997).

\bibitem{olive}
K.\ A.\ Olive and D.\ Thomas, hep-ph/9811444.

\bibitem{raffelt}
S.\ Hannestad and G.\ Raffelt, astro-ph/9805223

\bibitem{n}
A.\ Linde, Phys.\ Rev.\ D49, 748 (1994);
S.\ Mollerach, S.\ Matarrese and F.\ Lucchin, Phys.\ Rev.\ D50, 4835
(1994);
E.\ J.\ Copeland {\it et al.} Phys.\ Rev.\ D49, 6410 (1994).

\bibitem{tevap}
D.\ N.\ Page, Phys.\ Rev.\ D13, 198 (1976);
B.\ J.\ Carr, Astrophys.\ J.\ 206, 8 (1976).

\bibitem{carr}
B.\ J.\ Carr, Astrophys.\ J.\ 201, 1 (1975).

\bibitem{cobe}
C.\ L.\ Bennett, {\it et al.}, Astrophys.\ J.\ 464, L1 (1996).

\bibitem{nchdm}
A.\ R.\ Liddle, {\it et al.}, Mon.\ Not.\ R.\ Astron.\ Soc.\ 281, 531 (1996).

\bibitem{pt}
J.\ Garcia-Bellido, A.\ Linde and D.\ Woods, Phys.\ Rev.\ D54, 6040 (1996).

\bibitem{strings}
S.\ W.\ Hawking, Phys.\ Lett.\ B231, 237 (1989).

\bibitem{bubbles}
S.\ W.\ Hawking, I.\ G.\ Moss and J.\ M.\ Stewart, Phys.\ Rev.\ D26, 2681
(1882);
I.\ G.\ Moss, Phys.\ Rev.\ D50, 676 (1994);
H.\ Kodama, M.\ Sasaki and K.\ Sato, Prog.\ Theor.\ Phys.\ 68, 1979
(1982).

\bibitem{twobubble}
M.\ Y.\ Khlopov, R.\ V.\ Konoplich, S.\ G.\ Rubin and A.\ S.\ Sakharov,
hep-ph/9807343

\end{references}
\end{document}